\documentclass[letterpaper]{article} 
\usepackage{aaai25}  
\usepackage{times}  
\usepackage{helvet}  
\usepackage{courier}  
\usepackage[hyphens]{url}  
\usepackage{graphicx} 
\usepackage{natbib}  
\usepackage{caption} 
\usepackage{array}
\usepackage{booktabs}
\usepackage{algorithm}
\usepackage{algorithmic}
\usepackage{tikz}
\usetikzlibrary{shapes.geometric, arrows.meta, positioning}

\tikzstyle{process} = [rectangle, minimum width=3.8cm, minimum height=1.2cm, text centered, draw=black, fill=gray!10]
\tikzstyle{sideprocess} = [rectangle, minimum width=3.8cm, minimum height=1.2cm, text centered, draw=black, fill=blue!10]
\tikzstyle{arrow} = [thick,->,>=stealth]
\usepackage[capitalise,sort&compress]{cleveref}
\usepackage{newfloat}
\usepackage{listings}
\DeclareCaptionStyle{ruled}{labelfont=normalfont,labelsep=colon,strut=off} 
\floatstyle{ruled}
\newfloat{listing}{tb}{lst}{}
\floatname{listing}{Listing}
\title{An Adaptive Responsible AI Governance Framework for Decentralized Organizations}

\author {
    Kiana Jafari Meimandi\textsuperscript{\rm 1},
    Anka Reuel\textsuperscript{\rm 1},
    Gabriela Aranguiz-Dias\textsuperscript{\rm 1},
    Hatim Rahama\textsuperscript{\rm 2},
    Ala-Eddine Ayadi\textsuperscript{\rm 2},
    Xavier Boullier\textsuperscript{\rm 2},
    Jérémy Verdo\textsuperscript{\rm 2},
    Louis Montanie\textsuperscript{\rm 2},
    Mykel Kochenderfer\textsuperscript{\rm 1}
}
\affiliations {
    \textsuperscript{\rm 1} Stanford University\\
    \textsuperscript{\rm 2} LVMH\\
    corresponding author: kjafari@stanford.edu
}
\usepackage{bibentry}

\begin{document}

\maketitle

\begin{abstract}
This paper examines the assessment challenges of Responsible AI (RAI) governance efforts in globally decentralized organizations through a case study collaboration between a leading research university and a multinational enterprise. While there are many proposed frameworks for RAI, their application in complex organizational settings with distributed decision-making authority remains underexplored. Our RAI assessment, conducted across multiple business units and AI use cases, reveals four key patterns that shape RAI implementation: (1) complex interplay between group-level guidance and local interpretation, (2) challenges translating abstract principles into operational practices, (3) regional and functional variation in implementation approaches, and (4) inconsistent accountability in risk oversight. Based on these findings, we propose an Adaptive RAI Governance (ARGO) Framework that balances central coordination with local autonomy through three interdependent layers: shared foundation standards, central advisory resources, and contextual local implementation. We contribute insights from academic-industry collaboration for RAI assessments, highlighting the importance of modular governance approaches that accommodate organizational complexity while maintaining alignment with responsible AI principles. These lessons offer practical guidance for organizations navigating the transition from RAI principles to operational practice within decentralized structures.
\end{abstract}
%
\section{Introduction}
Artificial Intelligence (AI) has become a strategic driver of innovation across industries, prompting organizations to adopt AI systems at increasing scale and speed. Alongside this growth, calls for Responsible AI (RAI), AI that is fair, transparent, accountable, and aligned with ethical and legal norms, have moved from abstract principles to operational necessity~\cite{schiff2021explaining}. Regulatory initiatives such as the EU AI Act~\cite{diaz2023connecting}, the OECD AI Principles~\cite{yeung2020recommendation}, and standards like ISO/IEC 42001~\cite{dudley2024rise} reflect a growing consensus that AI development must be directed not just by technical performance but by societal values and institutional accountability.

While many frameworks and guidelines have emerged to support RAI, a critical implementation gap persists~\cite{reuel2024responsible}, especially for global organizations composed of independent or semi-independent business units that are functionally or geographically diverse. These enterprises face a complex landscape with regulatory inconsistencies across jurisdictions, cultural and operational diversity across regions, and limited centralized control over AI practices~\cite{al2025between}. Translating high-level RAI principles into action in such globally decentralized organizations remains underexplored in both research and practice.

This paper addresses that gap by presenting a case study of an RAI assessment collaboration between a group of researcher at a leading university and a global enterprise comprised of more than 50 business units, each operating with significant autonomy across various sectors. In 2024, an industrial collaborator initiated this assessment to inform the development of governance structures suitable for their decentralized organizational model, particularly focusing on how to balance centralized guidance with local adaptation. This objective aligned with the academic interest in studying implementation challenges in complex organizational settings, creating a foundation for meaningful collaboration.

Rather than developing a new RAI framework, our goal was to adapt existing frameworks across this diverse ecosystem. We sought to evaluate how well current RAI principles and tools translate into decentralized environments, what organizational and cultural barriers arise during assessment, and how collaboration between academia and industry can support meaningful evaluation and reflection. While the assessment was carried out as part of an academic partnership, our findings also translate to other third-party evaluation settings of decentralized companies.

The assessment specifically focused on internally developed AI systems rather than third-party solutions, addressing the unique challenges of a decentralized organization. We focus on four central questions:
\begin{itemize}
    \item How do group-level RAI guidelines interact with local implementation practices in globally decentralized organizations? 
    Aim: To explore a potential disparity between centralized guidance and independent interpretation across diverse business units and usage contexts.
    \item What specific mechanisms are implemented to translate abstract RAI principles into operational practices across diverse business contexts? Aim: To understand the practical challenges of moving from high-level values to concrete workflows and decision processes.
    \item How do regional and functional variations shape the interpretation and application of RAI frameworks within a multinational enterprise? Aim: To examine how cultural, regulatory, and operational diversity influences RAI implementation across geographies and business functions.
    \item What governance structures and accountability mechanisms effectively support RAI oversight in organizations with distributed decision-making authority? Aim: To investigate how responsibility in AI development and deployment can be appropriately assigned and enforced in complex organizational settings.
\end{itemize}
This paper contributes: (1) a detailed account of challenges encountered during a large-scale RAI assessment in a global, decentralized organization, (2) reflections on friction points between corporate standards and local realities, especially around enforcement, autonomy, and interpretation of principles, (3) practical insights into the logistics, politics, and methods of conducting RAI assessments at an enterprise scale, and (4) a critical look at the interface between academia and industry in operationalizing RAI practices.

By sharing lessons learned
we aim to inform both future research and practical governance efforts in similar organizational settings. The resulting Adaptive RAI Governance (ARGO) framework we propose offers a structured yet flexible approach to balancing centralized coordination with local autonomy in complex enterprises.

\section{Background}
As AI systems become central to organizational decision-making, concerns about their ethical, legal, and societal impacts have prompted the development of numerous RAI frameworks and guidelines~\cite{papagiannidis2025responsible}. These frameworks typically define principles such as fairness, transparency, accountability, and human oversight, and offer tools for assessing and mitigating risks across the AI lifecycle~\cite{schiff2021explaining}. While these efforts provide high-level guidance for AI governance, they often presume implementation conditions that are misaligned with the realities of globally decentralized organizations. This section examines how existing frameworks address, or fail to address, critical challenges of RAI implementation in complex organizational settings. 

\subsection{Governance and Accountability}
The ``Many Hand’’ problem~\cite{nissenbaum1996accountability} represents a fundamental challenge in AI governance; when responsibility is distributed across departments, geographies, and levels of autonomy, clear accountability becomes difficult to establish. Current RAI frameworks take varying approaches to this challenge. The NIST AI Risk Management Framework (AI RMF) acknowledges organizational complexity by emphasizing multi-stakeholder involvement and continuous improvement cycles~\cite{tabassi2023artificial}. However, it provides limited guidance on how to assign specific responsibilities across organizational boundaries.

ISO/IEC 42001:2023 takes a more structured approach, introducing a certifiable management system standard that positions RAI as a systems-level, auditable process~\cite{benraouane2024ai}. This standard offers clearer accountability structures but assumes an implementation capability that may not exist in organizations with highly decentralized decision-making. The diffusion of responsibility in complex organizations often undermines even well-designed accountability mechanisms~\cite{lo2020ethical}.

\subsection{Translating Principles to Practice Across Contexts}
A second challenge involves operationalizing abstract ethical principles in specific development contexts. The OECD AI Principles establish five high-level tenets: inclusive growth, human-centered values, transparency, robustness, and accountability~\cite{yeung2020recommendation}. While these principles have been adopted by over 40 countries, they offer limited operational guidance for implementation teams. Similarly, the IEEE Ethically Aligned Design (EAD) articulates detailed ethical guidelines but leaves substantial interpretive latitude for implementing teams~\cite{how2018ethically}.

This interpretive latitude leads to inconsistent implementation, as units within the same company may interpret principles like fairness or transparency differently based on regional context, (R)AI maturity \cite{reuel2024responsible}, or sectoral norms~\cite{morley2020initial}. This creates significant variation in practice, even when common corporate policies exist. Current frameworks rarely address the question of how much variation in interpretation is acceptable or how organizations should balance consistency with contextual appropriateness.

\subsection{Balancing Local Autonomy with Coordination}
Decentralized organizations face a particular challenge in balancing local operational autonomy with the need for central coordination of standards~\cite{knol2025strategic}. Several integration challenges have been identified when attempting to embed RAI into heterogeneous local processes, especially in organizations with decentralized IT, compliance, or model development structures~\cite{lu2024responsible}. Existing frameworks typically emphasize either centralized governance (as seen, e.g., in the EU AI Act's approach to high-risk systems) or local flexibility (as in principle-based approaches such as~\cite{corea2023principle}), but few address how to effectively combine these approaches~\cite{diaz2023connecting}.

This integration challenge affects how RAI considerations are prioritized against competing business objectives. Without clear incentives or internal accountability structures (e.g., audit processes or ethical review gates), RAI considerations may be given lower priority, especially in fast-paced product teams that often prioritize faster system or product deployment. Current frameworks offer limited guidance on creating balanced governance structures that respect operational autonomy while ensuring meaningful oversight.

\subsection{Managing Cross-Jurisdictional Regulatory Variation}
Global firms must comply with multiple AI governance regimes simultaneously, creating complex regulatory challenges. For example, the EU AI Act introduces legally binding requirements for ``high-risk’’ AI systems that are being placed on the European market, including documentation, human oversight, and conformity assessments~\cite{edwards2021eu}. the EU AI Act represents a shift from voluntary to regulatory governance. However, organizations operating across multiple jurisdictions must navigate these overlapping regulatory requirements alongside differing regional expectations.

This regulatory inconsistency complicates efforts to create uniform internal RAI protocols. A single AI system may need to satisfy different standards depending on where it operates, who it serves, or what use case it is being used for. Current frameworks typically acknowledge this challenge but provide limited practical guidance on designing governance systems that can efficiently address multiple regulatory regimes while maintaining operational coherence.

\subsection{RAI Implementation Gap}
Despite the depth of current RAI literature~\cite{batool2025ai}, there is limited empirical work on how these frameworks perform in organizationally complex and globally distributed contexts. Researchers have produced increasingly sophisticated metrics, documentation standards, and auditing tools~\cite{schiff2021explaining}, but these often fail to account for the practical constraints and organizational dynamics of large enterprises with decentralized structures.

This gap underscores the need for empirical studies that examine how organizations actually implement RAI across complex, multi-entity structures. Our research directly addresses this gap by studying the application of existing frameworks in a multinational corporation with highly independent business units, focusing on governance structures, implementation processes, and organizational adaptations that enable RAI practices across diverse operational contexts.

\subsection{Academic–Industry Collaboration on RAI}
Several recent efforts have called for deeper collaboration between academia and industry to close the gap in RAI between principle and practice~\cite{septiandri2024impact}. Three predominant collaborative models have emerged: knowledge transfer partnerships, where researchers develop tools later implemented by industry~\cite{raji2019actionable}; co-creation approaches, where frameworks are jointly developed by leveraging complementary expertise~\cite{holten2020shifting}; and action research, where researchers participate in organizational change processes while studying implementation dynamics~\cite{sloane2022participation}.

Despite these approaches, tools developed in academic settings often encounter resistance when they conflict with established workflows or business priorities~\cite{rakova2021responsible}.
While researchers have developed sophisticated measurement and auditing tools, these often ignore the practical realities of large organizations. 
Meanwhile, companies typically lack the time or expertise to implement these tools effectively without external help.

Collaborations between research institutions and industry create testing grounds for evaluating how well frameworks translate to applied settings. They enable mutual learning between researchers and industry partners: Researchers gain insight into institutional challenges and implementation realities, while industry partners benefit from structured state-of-the-art research and assessment methods as well ass external support and validation. However, existing research typically focuses on single organizational units or centralized governance, overlooking implementation challenges in decentralized, multi-unit organizations--a gap this research addresses through our collaborative assessment.

\begin{figure*}
    \centering
    \includegraphics[width=0.9\linewidth]{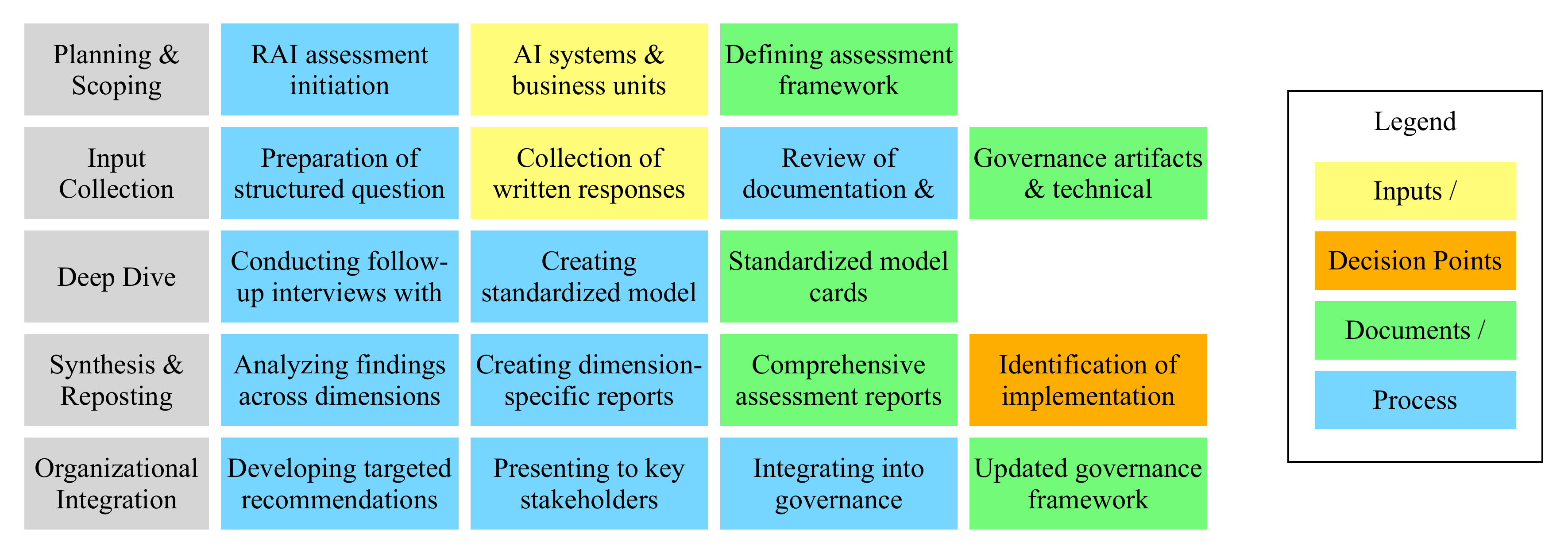}
    \caption{Responsible AI assessment process.}
    \label{fig:rai-assessment}
\end{figure*}
\section{RAI Assessment Context}
\subsection{Decentralized Global Enterprise}
The industry collaborator is a multinational corporation, composed of over 50 distinct business units operating across various industries such as retail, and hospitality. These business units function with a high degree of operational independence. Each maintains its own data infrastructure, development pipeline, and decision-making authority over the development and deployment of AI systems while still being officially affiliated with the enterprise and subsumed under and umbrella with the other business units.

This organizational structure presents specific governance challenges: AI systems are built and used in heterogeneous contexts, with varying levels of maturity, oversight, and regulatory exposure. An important organizational context worth noting is that more than half of the AI systems deployed across business units were co-developed at the enterprise level, creating both opportunities for centralized expertise and implementation testing, as well as challenges in adapting these systems to diverse business contexts.

At the corporate level, a group-wide RAI Charter has been issued, outlining high-level principles such as explainability, fairness, and privacy. While the Charter was formally applicable to all business units, which were accountable for building internal implementation mechanisms to be compliant, centralized compliance assessment mechanisms were still under development during the study period in 2024.


In this context, assessing the state of RAI practices requires attention not only to technical systems but also to the relationships between group-level structures and individual business units. The decentralized nature of this global enterprise thus provides a relevant case for examining the challenges of applying RAI frameworks across complex, multi-entity organizations.

\subsection{Collaboration Scope and Structure}
In 2024, the industry collaborators initiated a collaborative assessment effort with the academic collaborators at a research university to evaluate RAI practices across selected business units and use cases. The objective was to examine how existing RAI frameworks could be applied or adapted in a globally decentralized organizational setting, and to identify process-level and structural barriers to RAI implementation efforts.

The collaboration included the following components:
\begin{itemize}
    \item Framework alignment: Mapping the structure and content of existing RAI frameworks to the operational context of the global enterprise.
    \item RAI dimension-based assessment: Focusing on evaluating dimensions such as governance, transparency, and accountability across selected AI systems adapted from \cite{reuel2024responsible}.
    \item Stakeholder interviews: Conducting interviews with internal stakeholders involved in AI development, legal and compliance functions, and oversight roles.
    \item Documentation and process review: Analyzing available model documentation, AI development workflows, and AI governance materials at both the group and business unit level.
\end{itemize}

This structure allowed for a comparative and iterative process to explore how RAI-related principles were being interpreted and operationalized across varying contexts within the organization. The initiative was also intended to inform the gradual development of the organization's RAI Office and governance structures, including internal self-assessment processes and shared documentation practices.
\subsection{Scope of the Assessment}
The assessment was structured as a multi-phase process developed jointly by the industry and academic collaborators, with the aim of evaluating RAI practices across a representative set of algorithmic systems and organizational units~(\cref{fig:rai-assessment}). The approach combined both top-down (group-level governance) and bottom-up (project or use-case level implementation) perspectives, reflecting the decentralized nature of the organization’s structure.

Rather than pursuing a comprehensive audit, the focus was on identifying key challenges and gaps in translating RAI principles into practice within a complex organizational setting. The process included:
\begin{itemize}
    \item Mapping existing governance structures and documentation: The initial phase involved reviewing internal AI policies (such as the group-wide AI Charter), existing RAI-related roles, and development workflows, particularly as they varied across business units.
    \item AI use case  and portfolio scoping: AI use cases for assessment were selected through a scoping exercise conducted by the industry collaborators. Priority was given to frequently used or high-risk algorithms, with an emphasis on client-oriented or decision-critical use cases (e.g., product recommendation, sales forecasting, audience targeting). This process was organized in waves, beginning with a first batch of algorithms from selected business units.
    \item Dimension-based assessment: The assessment was organized around a set of predefined RAI dimensions: Diversity \& Fairness, Transparency, Privacy \& Data Governance, Human Interaction, Accountability, Reliability, Leadership \& Culture, and Societal and Environmental Wellbeing. Each dimension was explored through interviews, documentation review, and gap analysis.
    \item Stakeholder interviews: Interviews were conducted with individuals across group-level functions (e.g., legal, digital, AI governance) and unit-level teams (e.g., data scientists, product owners, project leads). The aim was to understand both strategic intentions and operational realities related to RAI.
    \item Iterative analysis and feedback: Insights were shared at key milestones, including internal presentations to stakeholders and contributions to organization’s internal events. The collaboration allowed for periodic refinement of both methodology and recommendations based on feedback.
\end{itemize}
The assessment was explicitly designed as a learning process, both to support internal reflection and to gather insights that help to adapt existing RAI frameworks for use in decentralized global enterprises.

The project plan spanned a one-year timeline (June 2024--May 2025), with structured checkpoints and thematic focus areas aligned to different calendar months. Rather than producing unidirectional, conclusive results, the assessment focused on context-aware observations and recommendations intended to inform internal strategy development, especially as the organization formalizes its RAI governance structures. This collaborative process also provided practical inputs that enabled the industry collaborator to iteratively define and implement key elements of their RAI strategy throughout the assessment.
\section{Methodology}
The evaluation framework covered nine dimensions of RAI, selected to align with major industry and regulatory frameworks based on the dimensions outlined in \cite{reuel2024responsible}, while also reflecting internal governance discussions. These dimensions were: Reliability, Privacy \& Data Governance, Diversity \& Fairness, Transparency, Human Interaction, Societal \& Environmental Wellbeing, Accountability, Compliance \& Lawfulness, Leadership, Principles \& Culture \cite{reuel2024responsible}. The assessment was carried out in waves, each focused on a selected group of algorithms used across different business units. Each wave involved gathering input from relevant stakeholders and investigating all nine RAI dimensions for each use case.

It is worth noting that at the time of collaboration, the industry partner was working with evolving definitions of these principles and dimensions rather than a fixed, shared taxonomy.

\subsection{Data Collection Methods}
Three primary data sources informed the assessment:

\textbf{Written responses and interviews.} The assessment began with semi-structured interviews, preceded by the distribution of question sets (see example questions in Appendix A) to key stakeholders. Midway through the process, this approach was revised. To increase efficiency and broaden participation, the team began requesting written responses to the assessment questions in advance, followed by targeted interviews to clarify and elaborate on specific points. This change enabled the inclusion of more contributors across functional and geographic boundaries and allowed for more contextualized follow-up discussions.

\textbf{Document review.} Supporting documentation, including AI development and deployment workflows, internal governance materials, and project-level documentation, was reviewed to cross-check reported practices and identify potential gaps. A standardized model card template, adapted from~\cite{mitchell2019model}, was used to guide the documentation of selected AI systems across the organization. A blank version of the model card is provided in the Appendix A.

\textbf{Internal tools and governance artifacts.} Where available, platform-level governance tooling and group-wide governance documentation (e.g., an RAI Charter, policy templates) were reviewed to understand the mechanisms in place to support or monitor RAI practices.
\subsection{Analysis Process}
For each dimension, findings were synthesized and documented in a dimension-specific report. These reports:
\begin{itemize}
    \item Summarized best practices from academic and industry literature,
    \item Assessed the current status of RAI-related practices within the organization,
    \item Identified strengths and implementation gaps,
    \item And included targeted suggestions for improvement
\end{itemize}
This structured reporting process ensured a consistent, repeatable evaluation format across dimensions and use cases. It also served to ground organization-specific recommendations in state-of-the-art RAI research and emerging regulatory expectations.
\subsection{Limitations}
The assessment was qualitative and formative in nature. It did not include direct access to system-level data or code, quantitative validation of model performance, fairness, or robustness, and application of automated RAI audit tools.

Findings were based on interviews, written responses, and supporting documentation and should be interpreted as insights into current implementation practices and challenges, rather than as conclusive assessments of system compliance or technical performance.
\section{RAI Implementation Patterns}
Through a cross-functional assessment of RAI practices across the large, globally distributed organization, we identified four recurrent areas where implementation approaches diverged. 
Each reveals important dynamics that affect how RAI principles are interpreted, applied, and maintained across business units operating with a degree of independence.
\begin{figure*}
    \centering
    \includegraphics[width=0.85\linewidth]{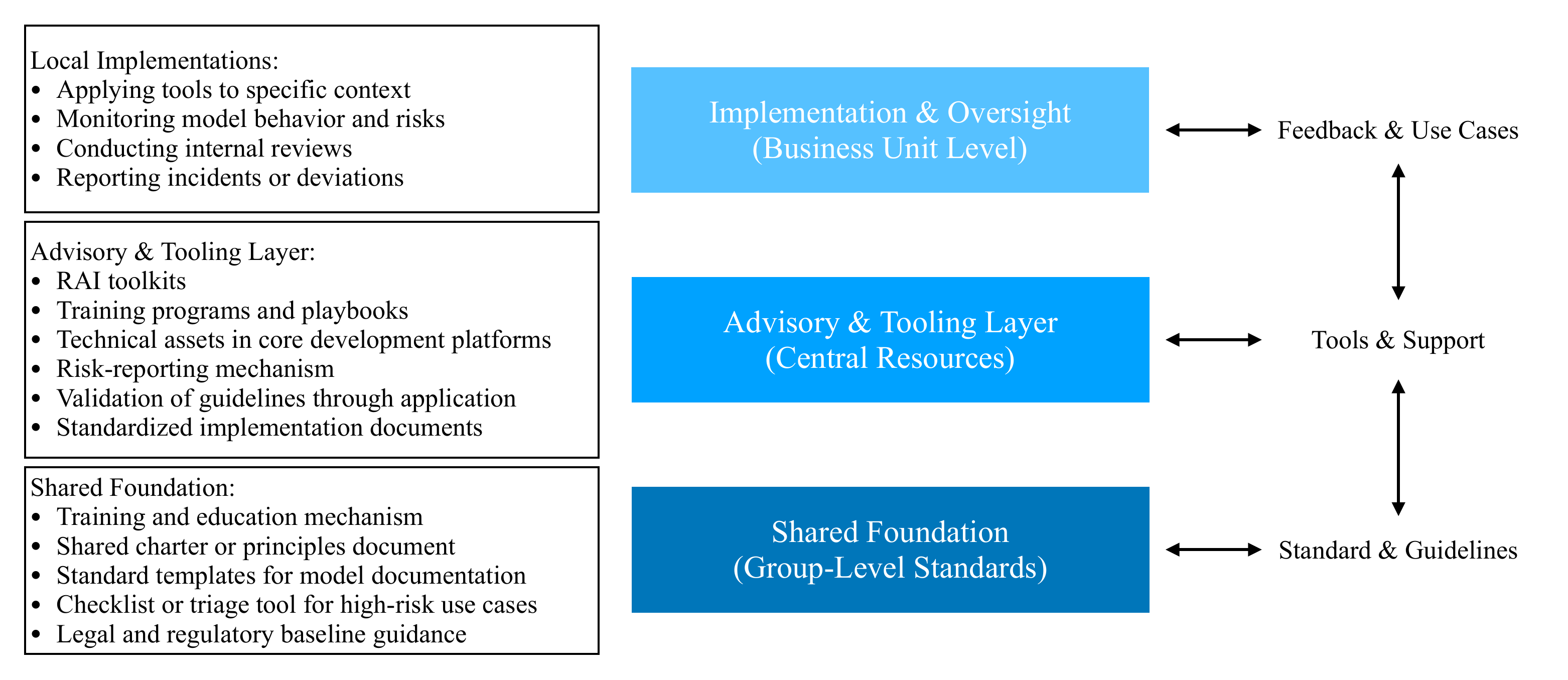}
    \caption{Adaptive RAI Governance (ARGO) Framework.}
    \label{fig:argo-framework}
\end{figure*}
\subsection{Group-Level Guidance and Local Implementation}
Organizations with a decentralized structure often define central principles or charters to guide RAI development at the group level. These high-level documents typically articulate shared commitments, and are intended to provide coherence across diverse business units.

However, our assessment revealed highly variable adoption of such guidance. Many group-level frameworks are advisory in nature, lacking sufficient implementation details or enforcement mechanisms. Business units are left to interpret and apply these principles according to their own priorities, AI maturity levels, and risk perceptions, with influence depending heavily on internal relationships and local RAI champions.

This results in variation in how (and whether) RAI practices are implemented at the operational level. While flexibility is respecting business units autonomy and local deployment contexts, it can also lead to uneven application of responsible safeguards and limited visibility into aggregated system-level risks.

\textit{Effective governance requires mechanisms that establish minimum shared standards while preserving space for contextual adaptation.}
\subsection{From Principles to Practice}
RAI frameworks often emphasize transparency, fairness, and accountability as guiding values. Yet, translating these into practical workflows, such as model documentation and risk assessment, remains challenging.

Technical teams have access to various tools for RAI implementation, including group-developed resources for explainability, fairness assessment, and environmental impact measurement. However, teams often struggled to connect these tools with high-level charter principles in ways that provided clear, actionable guidance for specific contexts.

Implementation varies widely across units. Some units incorporate fairness metrics into their model evaluation pipelines; others focus solely on performance indicators. Documentation quality and depth also vary significantly, influenced by team expertise, resource availability, and the perceived use case importance and riskiness. Teams expressed uncertainty about operationalizing principles in specific contexts-- questions like ``Which RAI metrics matter for this use case?'' often go unanswered without clear institutional guidance.

\textit{While frameworks and tools exist, their effective use depends on stronger integration into organizational processes and clearer mapping between abstract values and practical actions.}

\subsection{Regional and Functional Variation}
Large enterprises often operate across multiple regulatory environments and cultural contexts, introducing complexity into RAI implementation. 
Consent and disclosure expectations differ between jurisdictions, and some customer-based systems may be adjusted to comply with local norms inconsistently. Practices around demographic subgroup analysis, fairness testing, or localization of recommendations are not uniformly applied. 


This variation also appears within functional domains. AI applications in marketing, personalization, forecasting, or customer service are often developed using different assumptions and risk thresholds, even within the same organization. The extent to which RAI principles are operationalized depends on both geography, and the nature and perceived sensitivity of specific use cases.

\textit{Governance approaches must accommodate legitimate contextual variation while maintaining organizational alignment and shared accountability standards.}
\subsection{Roles and Responsibilities in RAI Oversight}
Oversight of AI development and deployment is typically distributed across strategy, risk, legal, and engineering functions. In decentralized organizations, these roles are often held across different teams, and business units, making responsibility for identifying and mitigating AI risks rarely centralized.

Risk identification and response processes remain fragmented. There is often no single escalation path for high-risk projects, and post-deployment incidents may be addressed within business units without feedback loops to the broader organization. Documentation and learnings from past challenges are inconsistently -- if at all -- shared across teams.

Some organizations have begun piloting self-assessment frameworks or peer review mechanisms to strengthen accountability~\cite{drach2024assessment}. However, these efforts are emergent and lack standardization, while the growth of generative AI and other high-impact technologies increase pressure to formalize roles, processes, and escalation protocols~\cite{leslie2024future}.

\textit{Clear role definition and escalation pathways are essential for distributed accountability, requiring formal mechanisms that connect local oversight with organizational learning.}

These four patterns point toward the need for governance approaches that can balance central coordination with local autonomy, translate abstract principles into actionable guidance, accommodate contextual variation while maintaining alignment, and distribute accountability in coordinated ways.

\section{Toward Adaptive Governance}

The four patterns identified in our assessment represent interconnected dimensions of a complex implementation ecosystem. Group-level guidance establishes the normative foundation that local implementation depends on. This relationship is bidirectional; local implementation experiences inform the evolution of central guidance over time if channels for communicating and sharing such learnings between business units and group-level entities exist.

The translation from principles to practice occurs at this intersection between group guidance and local implementation. How well this works depends on available resources, expertise, and organizational priorities in each setting. This translation process is also shaped by regional and functional variations, which introduce contextual factors that influence how abstract principles are interpreted in specific operational environments.

Accountability mechanisms span across these dimensions and can create feedback loops that either reinforce or weaken implementation consistency. When accountability is clearly assigned, it strengthens the connection between group-level guidance and local practices. When left underspecified, it widens the gap between principles and practices, allowing greater regional and functional divergence.

These patterns form a dynamic organizational system where changes in one dimension influence others. For example, strengthening role clarity in RAI oversight directly impacts how effectively principles translate to practice, while acknowledging regional variation can lead to more contextually appropriate central guidance.

Based on these findings, we propose an adaptive framework that addresses the governance gaps identified in our assessment. Unlike centralized models that create bottlenecks or fully decentralized approaches that result in inconsistent safeguards, our framework establishes a balanced governance structure combining central guidance with local adaptation. This three-layer approach directly responds to implementation challenges: a shared foundation layer addresses inconsistent interpretation of principles, an advisory layer bridges the gap between abstract values and operational practices, and a local implementation layer acknowledges contextual adaptation while maintaining organizational alignment.

\subsection{Adaptive Responsible Governance Framework}
Our proposed Adaptive RAI Governance (ARGO) Framework is designed to align central coordination with local autonomy, addressing the challenges observed in our study, particularly variability in implementation, unclear accountability, and inconsistent tool adoption, by offering a structured but adaptable approach. The ARGO Framework recognizes that AI development, deployment, and use occur across multiple organizational levels~(\cref{fig:argo-framework}). It defines three interdependent layers that work together to support consistent and contextually appropriate RAI practices.

\textbf{Shared Foundation (Group-Level Standards).}
The goal at this level is to define a  minimum set of expectations across the organization, including:
    \begin{itemize}
        \item A shared charter or document outlining RAI principles
        \item Standard templates for model and use case documentation
        \item A checklist or triage tool to identify high-risk use cases
        \item Legal and regulatory baseline guidance
        \item Clear definition of roles and responsibilities for RAI oversight at both group and business unit levels
    \end{itemize}
This layer addresses the need for consistent normative foundations while acknowledging that implementation will vary. By establishing minimum requirements rather than prescribing uniform processes, it respects the autonomy of business units while ensuring essential safeguards are in place, in particular for high-risk use cases, across all organizational units.
    
\textbf{Advisory and Tooling Layer (Central Resources).}
At this level, centralized group-level teams provide:
    \begin{itemize}
        \item RAI toolkits (e.g., fairness metrics libraries, explainability dashboards)
        \item Training programs and playbooks
        \item Technical assets integrated into core development platforms
        \item Lightweight reporting mechanisms to improve visibility across units
        \item Centralization of common AI assets at the group level with applied RAI practices to serve as implementation models
        \item Validation of guidelines through application to group-level AI assets
        \item Standardized templates and documentation guidelines for consistent implementation across business units
        \item Feedback channels for business units to report if any group-level assets are not working for their specific use case – or if they do particularly well.
    \end{itemize}
This layer addresses the need for consistent normative foundations while acknowledging that implementation will vary. By establishing minimum requirements and feedback processes rather than prescribing uniform processes in a unidirectional manner, it respects the autonomy of business units while ensuring essential RAI structures and safeguards are in place.

\textbf{Local Implementation and Oversight (Business Unit Level).}
    Individual teams or business units retain responsibility for:
    \begin{itemize}
        \item Applying tools and processes to their context
        \item Monitoring model behavior and risks
        \item Conducting internal reviews or self-assessments
        \item Reporting incidents or deviations
    \end{itemize}
This layer acknowledges the necessity of contextual adaptation while maintaining alignment with organizational commitments. By explicitly designating local responsibilities, it addresses the accountability challenges observed in our assessment.
\begin{table*}[h!]
    \centering
    \small
    \begin{tabular}{@{}>{\raggedright\arraybackslash}m{3.5cm} >{\raggedright\arraybackslash}m{4cm} > {\raggedright\arraybackslash}m{4cm}> {\raggedright\arraybackslash}m{4.5cm}@{}} 
        \toprule
        \textbf{Governance Dimension} & \textbf{Traditional Centralized Approach} & \textbf{Fully Decentralized Approach} & \textbf{ARGO Framework}\\
        \midrule
        \textbf{Decision Authority} & Concentrated in central ethics boards or committees & Dispersed across autonomous units with limited coordination & Layered authority with shared minimum standards and local contextualization\\
        \textbf{Implementation Guidance} & Prescriptive, uniform processes across the organization & Ad hoc, varying widely between units & Modular toolkits with core requirements and optional components\\
        \textbf{Regional/Cultural Adaptation} & Limited accommodation of regional differences & High variation with limited cross-regional learning & Explicit support for contextual implementation with shared learning mechanisms\\
        \textbf{Risk Assessment} & Standardized criteria applied uniformly & Inconsistent approaches based on local perceptions & Risk-based tiering with common assessment frameworks and local interpretation\\
        \textbf{Knowledge Sharing} & Formal channels with limited practitioner input & Fragmented knowledge with limited cross-unit transfer & Communities of practice and lightweight feedback mechanisms\\
        \bottomrule
    \end{tabular}
    \caption{Comparing the ARGO Framework against existing RAI governance models.}
    \label{tab:argo-comparison}
\end{table*}
\subsection{Implementation Recommendations}
Drawing on observed challenges and best practices, we offer the following design recommendations for multi-entity organizations seeking to implement or improve a decentralized RAI framework with central support.
\begin{itemize}
    \item \textbf{Define a core set of minimum practices} that all units must follow, regardless of risk level (e.g., model documentation, data provenance tracking).
    \item \textbf{Invest in shared, reusable tools} that make RAI efforts practical and efficient to implement, rather than burdensome or abstract.
    \item \textbf{Embed flexibility through modular toolkits}, allowing decentralized teams to select components based on their use case, maturity, and risk profile.
    \item \textbf{Create lightweight feedback and escalation mechanisms}, particularly for surfacing incidents, near misses, unexpected model behavior, or issues with group-level assets that arise at the business-unit level.
    \item \textbf{Prioritize visibility over control}, ensure the organization knows where AI is being used, how it is evaluated, and by whom, without overly prescriptive oversight.
    \item \textbf{Encourage shared learning}, such as cross-unit peer reviews or communities of practice, to prevent duplication of effort and accelerate ethical innovation.
\end{itemize}
\subsection{Positioning Relative to Existing Models}
The ARGO framework with central support builds on and complements emerging RAI standards. Unlike centralized governance models (e.g., where all risk review flows through a single ethics board \cite{schuett2024design}), it avoids bottlenecks by distributing ownership. Unlike fully decentralized models (where every unit develops its own practices), it ensures alignment through shared infrastructure.

Compared to ISO/IEC 42001 or the NIST RMF, this framework emphasizes organizational fit; integrating principles into existing workflows and allowing for gradual adoption across heterogeneous units. We illustrate the key differences between these approaches in~\cref{tab:argo-comparison}. 

The ARGO Framework is compatible with certifiable management systems but focuses more explicitly on cross-unit coordination, accountability layering, and sociotechnical alignment in complex organizational settings.
\subsection{Toward Standardization in Practice}
One of the key challenges in operationalizing RAI is ensuring that assessments are comparable across diverse projects and teams. The ARGO framework supports this by defining common inputs and outputs (e.g., standardized model cards or RAI metrics), encouraging consistent documentation across projects, and allowing different units to tailor implementation while still enabling portfolio-wide visibility and benchmarking. 

Ultimately, this enables leadership to assess the state of RAI maturity at both the unit and enterprise levels, without requiring full centralization of development or control.

The ARGO framework offers a practical path forward for global organizations managing AI across multiple dispersed entities. It acknowledges that RAI cannot be fully standardized nor entirely devolved. Instead, it must be layered; supported by shared infrastructure, guided by clear principles, and adapted locally through accountable implementation.

The next sections reflect on the academic–industry collaboration that informed this work, and offers targeted recommendations for both practitioners and researchers seeking to strengthen RAI practices in complex organizational settings.
\section{Academic–Industry Collaboration}
This project was structured as a collaborative assessment of RAI practices, conducted jointly by academic researchers and an enterprise organization operating across multiple business units. The collaboration aimed to produce actionable insights for the organization while contributing generalizable knowledge to the broader research and policy communities.

\subsection{Elements that Supported Collaboration}
\textbf{Complementary roles and knowledge.} The collaboration was informed by a shared interest in RAI, though each side contributed different forms of expertise. Academic researchers brought knowledge of state-of-the-art frameworks, standards, and broader field developments, and benchmarking insights from similar implementation efforts at other organizations. Industry participants provided access to organizational context, diverse use cases, and implementation details. This combination created a reciprocal dynamic that helped bring about questions and observations that might not have emerged from a single perspective.

\textbf{Structured process and ongoing coordination.} The collaboration followed a phased timeline with regular checkpoints. This structure helped accommodate different working styles and allowed for periodic reflection and adjustment. Meetings and milestones helped balance short-term needs with longer-term analytical goals.

\textbf{Joint development of insights.} Rather than applying an external framework in a static way, the process involved adapting tools and interpretations through discussion. Questions were refined based on organizational feedback, and existing practices were clarified in response to structured inquiry. This allowed both sides to contribute to shaping the analysis.
\subsection{Clarifying Motivations and Expectations}
An important observation from this project is that academic and industry participants often approach collaboration with different goals, constraints, and timeframes. Academic research typically emphasizes theoretical development, generalizable findings, and public dissemination. Timelines are often flexible, and researchers may not be positioned to provide direct operational recommendations. Industry participants, in contrast, may prioritize performance or operational improvement, regulatory compliance, or internal guidance. Timelines are often shorter, and information sharing may be subject to internal review processes or confidentiality restrictions.

These differences are not inherently problematic, but they can create misunderstandings if not discussed. 
Clarifying goals and expectations at the outset can help avoid such disconnects. This includes: \textit{discussing} what each side aims to gain, \textit{identifying} acceptable outputs, \textit{reviewing} data access and privacy limitations, and \textit{defining} what timelines and success indicators look like.

Another important consideration is how research findings are presented to internal stakeholders. Framing the assessment as a learning process rather than an audit significantly impacts stakeholder engagement and openness. When stakeholders perceive the process as evaluative (``what did you do wrong?'') rather than collaborative, defensive responses may limit meaningful exchange.

Having these conversations early and revisiting them periodically can help both sides structure the collaboration in a way that reflects their respective priorities and responsibilities.
\subsection{Considerations for Future Collaborations}
Structured collaborations between academia and industry provide mutual value. They support grounded interpretations of RAI frameworks and help organizations articulate internal practices. Researchers gain access to real-world assets and an understanding of operational constraints while generating approaches that balance theoretical rigor with practical considerations.

Partners should co-design collaboration structure and scope from the onset. Developing shared documentation tools that adapt across contexts proved valuable. Creating channels for structured reflection allowed both sides to discuss emerging insights and limitations. Most importantly, acknowledging different timelines and forms of value helped adapt communication accordingly.

Academia and industry partnerships on RAI require clear communication and shared understanding of roles, constraints, and expectations. When structured thoughtfully, such collaborations contribute to more robust and context-sensitive approaches to AI governance.
\section{Lessons Learned for Practice and Research}
Drawing on organizational assessment conducted across multiple semi-independent business units, this section synthesizes patterns observed in the implementation of RAI practices. These insights are presented as considerations for both practitioners working within complex enterprise structures and researchers studying or contributing to the development of RAI governance approaches.

The observations presented here are intended to inform future efforts by highlighting approaches that either frequently emerged or served as enabling conditions within the assessment context, rather than offering prescriptive guidance.
\subsection{Considerations for Practice}
\textbf{Contextualizing shared principles across units.} Several organizations maintain group-level charters or principles for AI governance that serve primarily as normative references rather than enforceable rules \cite{papagiannidis2025responsible}. These shared RAI commitments function as coordinating mechanisms, though their operationalization requires contextual adaptation. Local teams translate high-level principles differently depending on domain, region, and project maturity.

\textbf{Provision of non-prescriptive central resources.} Central teams frequently provide toolkits, documentation templates, and training materials without mandating their use. This non-prescriptive approach facilitates voluntary adoption while respecting local autonomy. Resources can include defined RAI metrics, model documentation structures, and internal review checklists. Centralization of such assets, as envisioned in the ``Advisory \& Tooling Layer'' of the ARGO Framework, supports broader internal alignment without rigid standardization.

\textbf{Integration of tooling into existing workflows.} Tool usability emerges as a critical factor in adoption. Resources embedded into platforms already used for model development or monitoring demonstrate higher likelihood of application in practice. This observation reinforces prior findings that practical integration matters more than policy mandates in enabling responsible practices during development.

\textbf{Clarification of roles and governance responsibilities.} Variability exists in how responsibility for RAI is distributed across business units and functions. In cases where governance responsibilities are explicitly defined~(e.g., with respect to documentation, review, or monitoring) implementation appears more consistent. The absence of role clarity often leads to ambiguity about whether, when, and how RAI processes should be applied.

\textbf{Lightweight feedback and documentation mechanisms.} Several teams implement informal feedback loops, such as sharing unexpected post-deployment model behavior or reflecting on incidents. While these mechanisms are often ad hoc, they support internal learning and normalization of ethical and responsible review. Structured but lightweight processes such as short model retrospectives or peer sharing of documentation serve as scalable entry points into more formal RAI governance practices.

\textbf{Focus on reusability and scalability.} Reusable templates and modular assessment tools demonstrate greater adaptability across units than customized or one-off resources. This aligns with a broader shift toward creating foundational infrastructure that can support variation in organizational structures and model types.

\subsection{Considerations for Research}
\textbf{Attention to distributed organizational structures.} Much of the existing literature on RAI governance focuses on centralized organizations or technology firms. The findings from this assessment suggest that distributed governance structures present distinct challenges and implementation patterns, including role ambiguity, inconsistent risk assessment practices, and varying local interpretations of shared principles. Future research can benefit from examining how governance operates across different business units, geographies, or product domains.

\textbf{Exploration of governance mechanisms beyond central review boards.} Organizational AI governance approaches may not rely on centralized review bodies. Instead, organizations employ self-assessments, team-led peer reviews, and embedded tooling. These practices raise questions about which governance models are feasible or effective in different contexts, and under what conditions more distributed accountability structures function effectively.

\textbf{Design of modular, context-aware tools.} Assessment tools that allow teams to select relevant components rather than apply a uniform checklist appear more compatible with local needs. Researchers developing toolkits or frameworks should consider how modularity and adaptability influence adoption, especially across teams with varying levels of RAI expertise.

\textbf{Value of co-designed evaluation approaches.} Collaborative development of evaluation processes between researchers and practitioners facilitates mutual understanding and local ownership. Co-design processes improve the relevance of academic frameworks and increase their applicability in operational environments. Further research can explore how co-design practices influence implementation outcomes in RAI settings.

\textbf{Support for standards translation rather than prescription.} While international standards (e.g., NIST, ISO/IEC 42001) offer valuable reference points, their application in decentralized organizations often involves adaptation. Researchers can contribute by studying how standards are interpreted or adjusted across sectors, and by clarifying which components are commonly operationalized.

\textbf{Interdisciplinary engagement.} The governance of AI in organizations intersects with legal compliance, systems engineering, social science, and computer science. Research efforts that integrate these perspectives are better positioned to identify friction points between ethical and responsible commitments and operational constraints

The observations presented in this section reflect the practices and governance dynamics in a large, distributed organization undergoing an RAI assessment. While not exhaustive or universally applicable, these lessons illustrate key considerations for designing governance approaches that balance structure and flexibility. They intend to inform ongoing research into how RAI principles translate into practice within diverse organizational contexts.

\section{Conclusion}
This study reveals that successful RAI implementation in globally decentralized organizations requires governance approaches that balance centralized guidance with local adaptation. Through our collaborative assessment across 50+ business units, we identified four key implementation patterns: the interplay between group-level guidance and local interpretation, challenges translating abstract principles into operational practices, regional and functional variation in approaches, and inconsistent accountability in risk oversight.

The proposed ARGO Framework addresses these challenges through three interdependent layers: shared foundation standards, central advisory resources, and contextual local implementation. This approach establishes minimum requirements while respecting operational autonomy. This offers a critical balance for enterprises operating across multiple jurisdictions and business contexts.

Our findings demonstrate that practical implementation often matters more than policy articulation. Tool adoption depends critically on integration into existing workflows, role clarity significantly impacts consistency, and modular resources enable adaptation across diverse contexts. The collaboration between academia and industry proved essential for generating insights that bridge theoretical frameworks with operational realities.

As AI capabilities advance and regulatory expectations increase, organizations need governance frameworks that can evolve. The ARGO approach offers a structured yet flexible path for ensuring AI systems align with societal values across complex organizational environments, contributing to the broader effort of closing the gap between RAI principles and practice.

\newpage
\bibliography{aaai25}

\clearpage
\appendix \label{app:A}
\section{Appendix A}
\subsection{Sample interview questions}
\textbf{Data and Privacy Questions}
\begin{itemize}
    \item Does your organization/your organization's business units have a formal data governance framework in place?
    \item Who is responsible for overseeing data governance, and how is accountability structured and enforced within the company?
    \item How does your organization/your organization's business units ensure that data used in AI systems is accurate, consistent, and up-to-date? Are there regular data updates?
    \item What processes are used to monitor and audit the lifecycle of data from collection to use and deletion?
    \item How are data quality issues, such as biased or missing data, identified and resolved?
    \item How are employees trained on data privacy practices?
    \item How does your organization/your organization's business units inform users or customers about their data rights and the use of their data in AI systems?
    \item Can users opt out of having their data used for AI model training, and how is this preference managed and honored?
    \item Are there channels for users to report concerns or request deletion of their personal data, and how is this process handled?
    \item Does your organization/your organization's business units have mechanisms for external stakeholders or regulators to review the company’s AI systems and data governance practices?
\end{itemize} 

\textbf{Data \& Third Parties}
\begin{itemize}
    \item What standards does your organization/your organization's business units impose on third-party vendors and partners when it comes to data privacy and governance?
    \item How are third-party vendors audited or monitored to ensure compliance with your organization's AI and data privacy standards?
    \item Does your organization/your organization's business units share or sell data with third parties, and what controls are in place to ensure that these transactions comply with privacy laws?
\end{itemize}

\textbf{Data Privacy \& Security}
\begin{itemize}
    \item Does your organization/your organization's business units use techniques like anonymization, pseudonymization, or differential privacy to protect individual/client data in AI systems? Can you describe how and what techniques you’re using in more detail?
    \item What processes are in place to detect and mitigate potential privacy breaches in AI systems?
    \item Are there policies for data retention and deletion? How is compliance with these policies enforced?
    \item Who is responsible for data privacy and how is accountability structured and enforced within the company?
    \item How does your organization/your organization's business units safeguard sensitive or personal data when used in training AI models?
    \item What measures are in place to ensure compliance with data privacy regulations such as GDPR?
    \item How does your organization/your organization's business units stay updated on evolving data privacy laws, regulations, and best practices that affect AI systems?
    \item Are there internal and/or external audits conducted regularly to ensure compliance with data privacy laws?
    \item What processes are in place for auditing AI models and ensuring that they remain compliant with privacy laws and organizational policies over time?
\end{itemize}

\textbf{Data Collection \& Usage}
\begin{itemize}
    \item To what extent does your organization/your organization's business units follow the principle of data minimization, collecting only the data that is necessary for AI applications?
    \item What types of data does your organization/your organization's business units collect to develop and deploy AI systems?
    \item How does your organization/your organization's business units ensure that the data used in AI models is legally and ethically collected?
    \begin{itemize}
        \item Ethically, in this context, means that the data is collected in a way that respects individuals' privacy, ensures informed consent, avoids bias or discrimination, and promotes fairness and transparency in how the data is gathered and used. This includes protecting sensitive information, not exploiting vulnerable groups, and adhering to standards of data integrity and accountability.
    \end{itemize}
    \item How is consent obtained from individuals whose data is used in AI models?
    \item Is the data obtained from third parties? If so, what verification process is in place to ensure compliance with privacy laws?
    \item Can individuals whose data is used in AI systems request information about how their data was utilized, and are they able to contest decisions made by AI systems, where the systems have impact on them?
    \item How does your organization/your organization's business units respond to privacy concerns or complaints raised by individuals impacted by AI systems? If this hasn’t happened before, is there a process in place in case an individual submits a complaint?
\end{itemize}
\subsection{Model Card}
\begin{table*}[h]
    \centering
    \small
    \begin{tabular}{>{\raggedright\arraybackslash}m{16cm} >{\raggedright\arraybackslash}m{0.05cm}} 
        \toprule
        \textbf{Basic Model Information} & \textbf{}\\
\midrule
        Organization or sub-entity developing model & \\
        Model date & \\
Model version & \\
Paper or other documentation for more information
\& Citation details (if applicable) & \\
License (if applicable) & \\
Model type \& reasoning: What type of model is it? Is it a generative or non-generative model? This also includes basic model architecture details, such as whether it is a Naive Bayes classifier, a Convolutional Neural Network, etc., and why this architecture was chosen. & \\
Description of input data & \\
Monitoring: Is model use monitored post-deployment (or is it planned)? If so, how? & \\
Contact person: Where to send questions or comments about the model & \\
\midrule        
\textbf{Intended Use Cases} & \\
\midrule
Primary intended uses (be as specific as possible) & \\
Primary intended users (also indicate here if it was meant to for internal use only or is it used by external parties, too) & \\
Out-of-scope use cases \& Limitations (e.g., not intended and tested for \\
\midrule
\textbf{Metrics} & \\
\midrule
Model evaluation measures: Any metric that is used to evaluate (aspects of) a model (e.g., performance, environmental impact, fairness, …) & \\
Decision thresholds: If decision thresholds are used, what are they, and why were those decision thresholds chosen? & \\
Approaches to measuring uncertainty and variability (if applicable) & \\
\midrule
\textbf{Evaluation Data} & \\
\midrule
Datasets: What datasets were used to evaluate the model?  & \\
Motivation: Why were these datasets chosen? & \\
Preprocessing: How was the data preprocessed for evaluation (e.g., tokenization  of sentences, cropping of images, any filtering such as dropping images without faces)?\\
\midrule
\textbf{Training Data} & \\
\midrule
Datasets: What datasets were used to train the model? What is the primary modelity of the data? How big was the dataset?\\
Motivation: Why were these datasets chosen?\\
Preprocessing: How was the data preprocessed for evaluation (e.g., tokenization of sentences, cropping of images, any filtering such as dropping images without faces)? \\
\midrule
\textbf{Ethical Considerations \& Mitigations} & \\
\midrule
Impacted Stakeholders: Beyond the direct users, who else might be impacted? Within the user base, are there subgroups that may be particularly vulnerable (e.g., due to their race, age, …, if applicable)? & \\
Data: Does the model use sensitive data (e.g., protected classes or PII)? If so, how is that handled? & \\	
Human life: Is the model intended to inform decisions about matters central to human life or flourishing – e.g., health or safety? Or could it be used in such a way?	& \\
Mitigations: What risk mitigation strategies were used during model development?	& \\
Risks and harms: What risks may be present in model usage? Try to identify the potential recipients, likelihood, and magnitude of harms. If these cannot be determined, note that they were considered but remain unknown.	& \\
Use cases: Are there any known model use cases that are especially fraught? This may connect directly to the intended use section of the model card.	& \\
Other concerns: This row should list additional concerns that were not covered. For example, did test results suggest any further testing? Were there any relevant groups that were not represented in the evaluation dataset? Are there additional recommendations for model use? & \\	
\bottomrule
    \end{tabular}
    \caption{This model card served as a starting point for the assessment based on~\cite{mitchell2019model}. We requested the assessed organization to provide model cards for all major AI models/systems used within the organization.}
    \label{tab:model-card}
\end{table*}


\end{document}